\documentclass[pdflatex,sn-mathphys-num]{sn-jnl}

\usepackage{graphicx}
\usepackage{multirow}
\usepackage{amsmath,amssymb,amsfonts}
\usepackage{amsthm}
\usepackage{mathrsfs}
\usepackage[title]{appendix}
\usepackage{xcolor}
\usepackage{textcomp}
\usepackage{manyfoot}
\usepackage{booktabs}
\usepackage{algorithm}
\usepackage{algorithmicx}
\usepackage{algpseudocode}
\usepackage{listings}
\usepackage{lipsum}
\usepackage[Lenny]{fncychap}
\usepackage{braket}
\usepackage{amsmath}
\usepackage{graphicx}
\graphicspath{{./figures/}}
\usepackage{caption}
\usepackage{subcaption}
\usepackage[normalem]{ulem}
\usepackage{comment}
\usepackage{amsthm}
\usepackage{hyperref}
\usepackage{amssymb}
\usepackage{stackengine}
\usepackage{bbm}
\usepackage{layouts} 
\usepackage{amssymb}

\begin{document}

\title[Channel capacity of small modular quantum networks in the
ultrastrongly coupled regime]{Channel capacity of small modular quantum networks in the
ultrastrongly coupled regime}

\author*[1]{\fnm{Salvatore Alex} \sur{Cordovana}}\email{salvocordovana99@gmail.com}
\author[1,2]{\fnm{Luigi} \sur{Giannelli}}
\author[1]{\fnm{Nicola} \sur{Macrì}}
\author[3,4]{\fnm{Giuliano} \sur{Benenti}}
\author[1,2,5]{\fnm{Elisabetta} \sur{Paladino}}
\author[1,2]{\fnm{Giuseppe A.} \sur{Falci}}\email{giuseppe.falci@unict.it}

\affil*[1]{\orgdiv{Dipartimento di Fisica e Astronomia "Ettore Majorana"}, \orgname{Università di Catania}, \orgaddress{\street{Via Santa Sofia 64}, \city{Catania}, \postcode{95123}, \state{Italy}}}

\affil*[2]{\orgdiv{Istituto Nazionale di Fisica Nucleare}, \orgname{Sezione di Catania}, \orgaddress{\street{Via Santa Sofia 64}, \city{Catania}, \postcode{95123}, \state{Italy}}}

\affil*[3]{\orgdiv{Center for
 Nonlinear and Complex Systems, Dipartimento di Scienza e Alta Tecnologia}, \orgname{Universitá degli Studi dell’Insubria}, \orgaddress{\street{Via Valleggio 11}, \city{Como}, \postcode{22100}, \state{Italy}}}
 
\affil*[4]{\orgdiv{Istituto Nazionale di Fisica Nucleare}, \orgname{Sezione di Milano}, \orgaddress{\street{Via Celoria 16}, \city{Milano}, \postcode{20133}, \state{Italy}}}

\affil*[5]{\orgdiv{CNR-IMM}, \orgname{Catania (University Unit), Consiglio Nazionale delle Ricerche}, \orgaddress{\street{Via Santa Sofia 64}, \city{Catania}, \postcode{95123}, \state{Italy}}}

\abstract{
We investigate state-transfer in modular quantum computer architectures exploiting the ultrastrong coupling regime of interaction between quantum processing units and   ICs. We show that protocols based on 
adiabatic coherent transport may achieve near-ideal single-letter quantum capacity and robustness against parametric fluctuations suppressing leakage induced by the dynamical Casimir effect.}


\maketitle

\section{Introduction}\label{sec1}
Developing modular architectures is recently attracting a large interest as a promising roadmap for upscaling quantum computers\footnote{See \href{https://www.ibm.com/quantum/blog/ibm-quantum-roadmap-2025}{https://www.ibm.com/quantum/blog/ibm-quantum-roadmap-2025}.} on solid-state platforms,   
since problems related to correlated decoherence, crosstalk, reliability of the subunits, complexity of control and of power and cooling infrastructures 
may be substantially softened. 

Modular hardware integrates interconnected quantum processing units (QPUs) onto a customized circuit enabling precise tuning and control of a dense architecture of qubits~\cite{warren_long-distance_2019,niu_low-loss_2023}. Interconnects (ICs) carry quantum information, therefore they allow quantum communication at the intercore level. State-transfer and swapping, entanglement sharing or multipartite entanglement generation are some of the main tasks to be performed.
They must be operated by local control of the QPUs and by switching their interaction with the ICs. 

We study simple models of quantum ICs to address state transfer between QPUs with sufficient immunity from errors to allow intercore fault tolerant computation. Sending states along a quantum channel is a basic task of quantum communication~\cite{Benenti,schumacher}. Leveraging this analogy we estimate the quantum capacity $\cal Q$ which is the figure of merit for efficient state-transmission according to the noisy quantum channel theorem~\cite{schumacher}.

A key point is the speed of quantum intercore operations and whether it can be comparable to intracore operations. Simple arguments suggest an estimate of the minimum clock time scaling as $T_{op} \sim 1/g$ where $g$ is the QPUs-IC interaction. However, faster operations have a cost in terms of fidelity and in practical cases intercore operations are much slower than intracore ones~\cite{warren_long-distance_2019,niu_low-loss_2023}. We will address the tradeoff between speed and fidelity in the simple case where the QPUs are single qubits, $\mathrm Q_1$ and $\mathrm Q_2$. We model the  IC  by a $d \ge 2$ node of equally-spaced energy levels studying how our results scale with $d$ towards the limit $d \to \infty$ of the  IC  being a quantized harmonic mode.

If the $\mathrm Q_i$-IC interaction conserves the number of excitations ${\cal N}$, the dynamics under parametric driving is confined to ${\cal N}=0,1,$ subspace. Then state transfer can be operated with no error, for instance  by switching on and off in a sequence the two couplings to produce two state-swaps this protocol achieving the ideal scaling $T_{op} = 2 \pi/g$.
However, for increasing values of $g$ realistic interactions do not conserve $\cal N$ but possibly its parity. As a consequence, parametric driving creates pair of excitations determining leakage from the ${\cal N}=0,1$ subspace which deteriorates the quantum operation. In the $d \to \infty$ limit this phenomenon is usually referred as the dynamical Casimir effect (DCE). 
 
In this work we investigate state-transfer by two protocols, namely the quantum bus (QB) protocol consisting of two Rabi state swaps $Q_1 \to IC $ and $IC  \to Q_2$ operated by switching on sequentially the respective interactions and a protocol inspired to coherent transport by adiabatic passage (CTAP)~\cite{article:spatial_adiabatic_passage}. Ideally, CTAP may yield state-transfer never occupying the IC thus operating as a {\em virtual} quantum bus. This should limit DCE-induced errors also for fast operations performed in the ultrastrong coupling regime, i.e. up to values of $g$ comparable with the natural frequencies $\omega_c$ of the subunits of the network. Our results indicate that state transfer with ${\cal Q} \sim 1$ is achieved which is moreover robust against parametric fluctuations. 

\section{Model and protocols}
\begin{figure}[t!]
    \centering
    \includegraphics[width=0.9\textwidth]{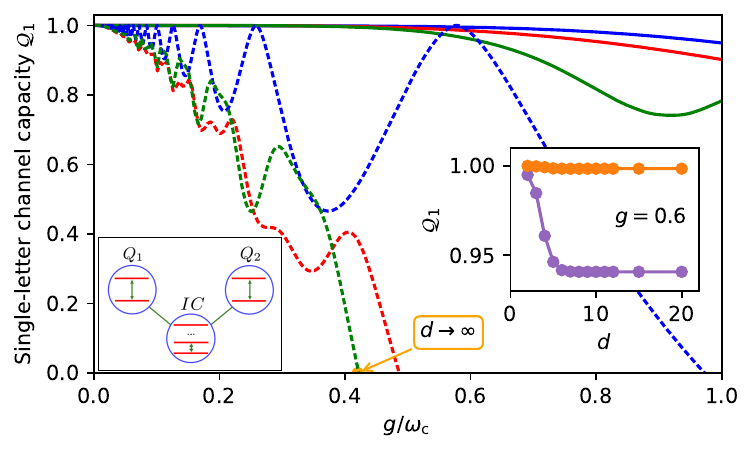}
    \caption{Single-letter channel capacity $\mathcal{Q}_1$ 
    vs $g$  for the QB (dashed lines) and for the CTAP (solid lines) protocols  for IC with $d=2$ (blue), $d=3$ (red) and  $d=4$ (green). 
    If the IC is a harmonic oscillator $\mathcal{Q}_1$ for the QB vanishes at $g\approx 0.42$ (orange dot)~\cite{article_benenti}. 
    CTAP is operated with Gaussian pulses with $gT =20$ and $\tau = 0.7\, T$ from the initial $t_i=-2 \sqrt{2}\,T$ to the final $t_f= 2  \sqrt{2}\,T$. 
    Left-bottom inset: schematics of the system consisting of two qubits coupled to a $d$-level interconnect. Right inset: $\mathcal{Q}_1$ as function of $d$ for CTAP at $g=0.6$ for $gT=20$ (purple) and $gT=40$ (orange) saturating in both cases to a constant value for increasing $d$.
    \\
    }
    \label{fig:Q1_protocols}
\end{figure}

We consider a quantum network composed by two qubits interacting with an IC 
modeled by a $d$-level system (see inset of Fig.\ref{fig:Q1_protocols}) with 
Hamiltonian ($\hbar=1$)
\begin{align}\label{eq:system_Hamiltonian}
& H(t)= P_d \,\big[\omega_c\,a^\dagger a + \sum_{i=1}^{2} \big( \epsilon_i\,\sigma_i^+\sigma_i^- + g_i\,f_i(t) \, (a^\dagger + a )\, (\sigma^+_i + \sigma^-_i) \big) \big]\,P_d 
\end{align}
where the qubits  $i=1,2$ have splittings $\epsilon_i$,  the usual ladder SU(2) operators are $\sigma_i^{\alpha}$
for $\alpha = \pm$, the  IC  has splitting $\omega_c$, field operators $a$ and $a^\dagger$ and $P_d$ projects onto its $d$ lowest levels.  

State-tansfer is achieved by modulating the qubits-IC  couplings $g_if_i(t)$. This Hamiltonian does not conserve the number of excitations, ${\cal N} = P_d \, a^\dagger a\, P_d  + \sum_i \sigma_i^+\sigma_i^- $, but only its parity, $\Pi=e^{i\pi {\cal N}}$. An ${\cal N}$-conserving Hamiltonian is obtained in the rotating-wave approximation (RWA) where 
the "counter-rotating" terms 
$a\sigma_i^-+a^\dagger\sigma_i^+$ are ignored. 
In the RWA perfect operations can be performed by choosing suitable $f_i(t)$.  
The presence of counter-rotating terms deteriorates the fidelity due to DCE~\cite{article_benenti,article_Stramacchia}. 

For state-transfer, qubit 1 is prepared in the state $\rho$ whereas qubit 2 and the  IC 
are initialized in their respective ground states, $\ket{0}_{Q_2}$ and $\ket{0}_{IC}$. We are interested to the final state of qubit 2, given by the quantum map 
\begin{equation}\label{map_rho}
    \rho^\prime = \mathcal{E}(\rho)=\text{Tr}_{Q_1, IC} \left[ U(t) \left( \rho \otimes \ket{0}\!\bra{0}_{IC} \otimes \ket{0}\!\bra{0}_{Q_2} \right) U^\dagger(t) \right],
\end{equation}
where $U(t)$ is the evolution operator of the whole system. 
The theory of quantum communication assesses the performance of the channel $\mathcal{E}$ by the \textit{quantum capacity} $\mathcal{Q} \le 1$ which quantifies the efficiency in transmitting qubits reliably, meaning that the errors induced by the channel can be  corrected by a quantum protocol~\cite{Benenti,Giovannetti2024}.
For memoryless channels~\cite{schumacher,Benenti}, it is defined by considering $N$ uses of the channel, then maximizing the \textit{coherent information} $I_c$ over all possible $N$-qubit input states $\rho_N$ and finally, by taking the $N\to \infty$ limit,  ${\cal Q} = \lim_{N \to \infty} {\mathcal Q_N}/{N}$, where
\begin{equation}
\mathcal{Q}_N = \max_{\rho_{N}} I_c(\mathcal{E}_N, \rho_N) \;,
 \qquad   I_c(\mathcal{E}_N, \rho_N) = S[\mathcal{E}_N (\rho_N)] - S_e(\mathcal{E}_N, \rho_N) \quad.
\end{equation}
Here $\mathcal{E}_N$ is the superoperator describing $N$ uses of the channel which reduces to $\mathcal{E}^{\otimes N}$ in the memoryless case, $S(\rho) := -\mathrm{Tr}[\rho \log_2 \rho]$ is the von Neumann entropy and $S_e(\mathcal{E}_N, \rho_N) := S\left[(\mathbbm{1}_N \otimes \mathcal{E}_N)(|\psi_N \rangle \langle \psi_N|)\right]$ is the \textit{entropy exchange}, $\ket{\psi_N}$ being any purification of \( \rho_N \).

In this work, we make two simplifying assumptions. First, we estimate 
$\mathcal{Q}$ by evaluate the \textit{single-letter channel capacity} $\mathcal{Q}_1$, this choice being discussed in the conclusions. Second, we assume that $I_c(\mathcal{E}, \rho)$ is maximized by the unpolarized $\rho= {1 \over 2} \mathbbm{1}_1$. This is true for both for the QB and the CTAP protocols
if the Hamiltonian is in the RWA. We assume that the error in the maximization of
the full Hamiltonian Eq.(\ref{eq:system_Hamiltonian}) is negligible, as it has been verified in Ref~\cite{article_benenti} for the QB protocol.

For simplicity, we consider state-transfer protocols operated at resonance, 
$\epsilon_1 = \epsilon_2 = \omega_c$, and by symmetric couplings, $g_1 = g_2 =:g$. Then the two swaps $Q_1 \to IC$ and $IC \to Q_2$ are operated  by
switching on sequentially the corresponding interaction for a time ${\pi}/({2g})=: T_{op}/2$, i.e.  by taking rectangular $f_i(t)$ in Eq.(\ref{eq:system_Hamiltonian}).
Instead, in the CTAP-inspired protocol
\cite{article:shreyasi_CTAP, article:luigi_tutorial_STIRAP,reference_CTAP_1,gullans_coherent_2020} qubits are coupled to the IC via
\begin{equation}
    f_1(t) = F[(t-\tau)/T] ,\quad
f_2(t) = F[(t+\tau)/T]
\end{equation}
where $F(t)$ is a slowly varying function with width $\sim T$ and $\tau > 0$  the half delay implying that the IC is first coupled to Q$_2$ and then to Q$_1$ implementing the so-called "counterintuitive sequence".

In the RWA, CTAP conserves ${\cal N}$ and may yield complete population transfer $\mathrm Q_1 \to \mathrm Q_2$  by adiabatically transferring the single-excitation along a "trapped state" of the two-qubits subspace, the IC being never populated. This requires sufficient adiabaticity, quantified by the well-known "global adiabiticity condition"~\cite{article:luigi_tutorial_STIRAP}, $g T > 10$. 
Larger values of $gT$ suppress transitions between instantaneous eigenstates yielding larger efficiency, as it is apparent in the inset of Fig.~\ref{fig:Q1_protocols}. 
The protocol is  robust against parametric fluctuations except for sensitivity to a non-zero "qubit detuning" $\epsilon_1-\epsilon_2$. By operating on a superposition $\ket{\psi}= c_0 \ket{0} + c_1 \ket{1}$ both the ideal QB (within a global phase factor) and CTAP transfer the state $\ket{\psi}_{Q_1} \otimes \ket{0}_{IC} \otimes \ket{0}_{Q_2} \to \ket{0}_{Q_1} \otimes \ket{0}_{IC} \otimes \ket{\psi}_{Q_2}$. 

By increasing $g$ counterrotating terms become important and state-transfer will suffer from DCE-induced leakage from the ${\cal N }=0,1$ subspace. Moreover, for not too large $g$'s the low-energy effective Hamiltonian is renormalized due to dressing by virtual photons which may affect the efficiency of the CTAP protocol. 

\section{Results}
Results for the single-letter channel capacity $\mathcal{Q}_1$ as function of $g$ are presented in Fig~\ref{fig:Q1_protocols}. 
For the  QB protocol (dashed lines) $\mathcal{Q}_1$ shows an oscillating behavior and it is suppressed as long as the number $d$ of states of the IC increases, which is a natural instance for a realistic model. The case $d=2$ can be calculated in closed form while the limit $d \to \infty$ was studied in Ref.~\cite{article_benenti} showing that ${\cal Q}_1 \ge 0$ for $g \,\lesssim 0.42$ properly shaped pulses reducing the amplitude of the oscillations.

The CTAP protocol (Fig.~\ref{fig:Q1_protocols} solid lines) shows remarkably larger values of $\mathcal{Q}_1$ with respect to the QB case for intermediate $0.1 \le g \le 1$, i.e. well inside the ultrastrong coupling regime. The results were obtained using Gaussian pulses $F(x) = \mathrm e^{-x^2}$. Notice that $\mathcal{Q}_1$ does not display oscillatory behavior implying that the protocol is robust against fluctuations of $g$ and of $T$. Actually, CTAP is well known to be also almost insensitive to the pulse shape $F(x)$ and robust against fluctuations of other parameters as $\tau$ or the detuning $\omega_c-\epsilon_i$ being instead sensitive to fluctuations of the qubit detuning, $\epsilon_1 - \epsilon_2$~\cite{bergmann_roadmap_2019}. Notice that the detunings can be set by local operations modifying the splittings $\epsilon_i$. 
The inset Fig.~\ref{fig:Q1_protocols} shows how $\mathcal{Q}_1$ scales with $d$ for the CTAP protocol at intermediate couplings in the ultrastrong regime. For increasing $d$, the capacitance quickly reaches a constant value, suggesting that CTAP only explores the lowest-energy levels of the interconnect. 

\begin{figure}[t!]
    \centering
    \includegraphics[width=0.9\textwidth]{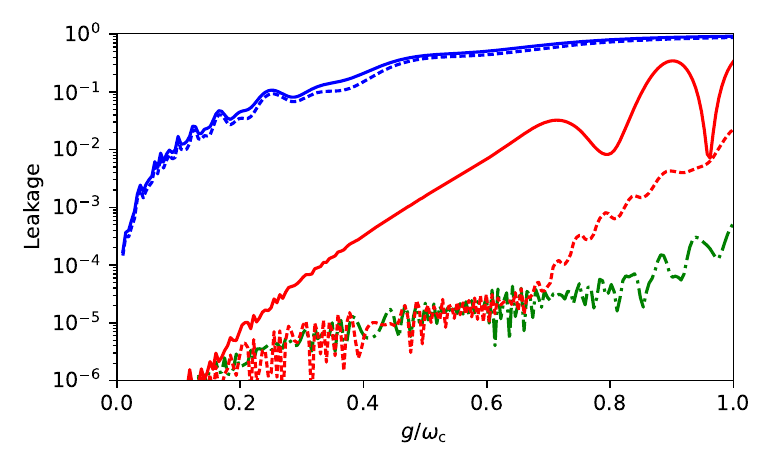}
    
    \caption{Leakage from the target subsystem $\{\ket{0}_{Q_1} \otimes \ket{0}_{IC} \otimes \ket{\psi}_{Q_2} \}$ (solid lines) and from the subspace ${\cal N}=0,1$ excitations (dashed lines and dot-dashed lines), for the QB (blue lines) and the CTAP (red lines) protocols. Parameters are the same as in the previous figure. The numerical error is $<5\cdot10^{-5}$. All the curves refer to an IC with $d > 8$ equispaced levels excepect the dot-dashed line referring to $d=4$. This suggests that anharmonicity of the IC may suppress DCE-induced errors in the CTAP protocol.}
    \label{fig:Leakage}
\end{figure}

Further insight in the physics of state-transfer is gained by studying the leakage at the end of the protocol from relevant subspaces of the Hilbert space shown in Fig.~\ref{fig:Leakage} for both QB and CTAP.
The leakage from the low-energy subspace ${\cal N}=0,1$ (dashed and dot-dashed curves) quantifies DCE-induced creation of pairs of excitations showing that CTAP suppresses pair production by four orders of magnitude with respect to the QB protocol up to values of $g < 0.7 \omega_c$. 
For larger values of $g$, leakage for CTAP depends on $d$ suggesting that it may be reduced by anharmonicity also in this regime or alternatively by increasing adiabaticity (see the inset of Fig.~\ref{fig:Q1_protocols}). Another interesting quantity is leakage from the target subsystem (solid lines), i.e. the subspace $\{\ket{0}_{Q_1} \otimes \ket{0}_{IC} \otimes \ket{\psi}_{Q_2} \}$, which suffers from an additional error induced by virtual-photon dressing which produces parametric imperfections renormalizing the low-energy Hamiltonian.
This error can be significant for CTAP becoming very large for $g > 0.7 \omega_c$. However, such errors may be correctable by more advanced control schemes using modulation of the local splittings $\epsilon_i(t)$ this topic being outside the scope of the present work. 
Notice that for the simple QB protocol, the error is almost entirely leakage from the subspace, due to the parametric modulation of the Hamiltonian producing pairs of excitations, as it happens in the DCE.

\section{Conclusions and open questions}\label{sec_conclusions}
In this work, we characterize a communication channel modeling an on-chip quantum IC between two QPUs. While high-speed intercore operations in a quantum network could be achieved in the regime of ultrastrong coupling between its subunits, this has a cost in terms of fidelity. In particular, for circuit QED-like distributed architectures a fundamental limit comes from the DCE. We studied two state-transfer protocols the IC being used as a quantum bus for real or virtual excitation. In this latter case, we found that the single-letter quantum channel capacitance ${\cal Q}_1$ is nearly unitary up to large $g \lesssim 0. 6 \,\omega_c$ and it shows a remarkable dependence on the non-linearity of the IC. We also studied DCE-induced leakage from the subspace ${\cal N}=0,1$ and leakage from the target subspace which shows additional  errors due to renormalization of the low-energy effective Hamiltonian by the counterrotating interaction. 

In quantum communication, the quantum capacitance ${\cal Q}$ is a rigorous measure of the ability to reconstruct a quantum letter from an $N$-letter use of a Markovian channel. In modular computing the scenario is different since we cannot send a large number $N$ of copies of the state. We used ${\cal Q}_1$ as a quantifier since it may provide at an heuristic figure of merit for a protocol which is not indefinitely repeated. For the same reason, the transfer rate defined as the capacity divided by $T$ is not a proper figure of merit of the IC. 

Aiming at applications to modular quantum computing, we mention that CTAP is an actual subject of investigation on solid-state platforms~\cite{reference_CTAP_1}. The proposal can be reformulated in terms of a related protocol called STIRAP~\cite{bergmann_roadmap_2019}, which has been proposed~\cite{di_stefano_coherent_2016,vepsalainen_quantum_2016} and demonstrated~\cite{kumar_stimulated_2016} for performing quantum operations in superconducting architectures.
We also stress that while the QB protocol requires hardware switchable on very short time scales, $T_{sw} \ll \pi/g$, CTAP and STIRAP are much less demanding,  $T_{sw} \ll 20 /g$. Implementations of these protocols in have been proposed with reduced available control~\cite{di_stefano_population_2015,brown_reinforcement_2021} in different drive configurations~\cite{pope_coherent_2019} and in the ultrastrong coupling regime~\cite{falci_advances_2017,falci_ultrastrong_2019,giannelli_detecting_2024}. 

We finally observe that since the IC can be left in an excited state the channel in general has memory for repeated uses. However, we expect such memory effects to be less important for CTAP where the IC is not populated which justifies the memoryless formalism used in this work.

\bmhead{Acknowledgements}
EP and LG acknowlwdge support by the PNRR MUR project PE0000023-NQSTI.
GF is supported from the ICSC – Centro Nazionale di Ricerca in High-Performance Computing, Big Data and Quantum Computing.
NM is supported from the PRIN 2022WKCJRT progect  SuperNISQ.
GF and EP acknowledge support from the University of Catania, Piano Incentivi Ricerca di Ateneo 2024-26, project QTCM. EP acknowledges the COST Action SUPERQUMAP (CA 21144).

\noindent
\bf Data Availability Statement\rm. This manuscript has no associated data or the data will not be deposited. [Authors’ comment: All the numerical data in this publication are available on request to the authors.]

\noindent
\bf Competing Interests\rm. The authors declare they have no financial interests nor competing interest to this work.

\bibliography{sn-bibliography}

\end{document}